\documentstyle[preprint,aps]{revtex}
\def\be{\begin{equation}}
\def\ee{\end{equation}}
\def\bea{\begin{eqnarray}}
\def\eea{\end{eqnarray}}
\def\ba{\begin{array}}
\def\ea{\end{array}}

\def\sq{\sqrt}
\def\al{\alpha}
\def\bt{\beta}
\def\ga{\gamma}

\def\ka{\kappa}

\def\om{\omega}

\def\ra{\rangle}

\begin{document}
\bibliographystyle{prsty}
\draft

\title{\bf New vacuum state and symmetry breaking in polariton system }
\author{Kelin Wang, Jinlong Yang, Tongzhong Li and Gan Qin}
\address
    {Center for Fundamental Physics,\\
     University of Science and Technology of China\\
     Hefei, Anhui, 230026, P. R. China}
\date{\today}
\maketitle
\begin{abstract}
The polariton system is studied by a concise approach using a simple model.
A new ground state with negative energy is obtained and found to 
exhibit the symmetry breaking. 
\end{abstract}

\vskip 0.5cm
\pacs{PACS numbers: 71.36.+c, 11.30.Qc\\
      {\bf Key Words}: Polariton, Symmetry Breaking}

\narrowtext

   The nonclassical properties of light have become a fascinating subject
of widespread investigations in recent years due to their importance in
the research of laser and the design of new light sources. Some
nonclassical states of light such as the squeezed states, the
pair-coherent states, and the photon number states have been extensively
studied both theoretically and experimentally\cite{Wall,Agar,Loud,Wu,Raiz,Mach}.
Recently, efforts have been made to study the nonclassical behavior of light in
model solid state systems\cite{Shar1,Shar2}. When light falls on a solid-state
material and interacts with the vibrating lattice, a photon-phonon complex can
be formed, which is called a polariton. Ghoshal and Chatterjee\cite{Shar1,Shar2}
have considered two possible model polariton systems, which involve one mode
of the photon field interacting with a single optical phonon, and solved them
by a canonical transformation to the Hamiltonian. Their results show that both
the phonon and photon subsystems can exhibit nonclassical behavior.

    In this Letter, we report the ground-state properties of the polariton
system. We adopted the model of Ghoshal and Chatterjee\cite{Shar2} for the
polariton system, and used a concise approach\cite{Wang}, where the wave
function is taken as the coherent-like form, to solve the Schr\"{o}dinger
equation exactly. We obtained a new ground state whose energy is lower than
the ordinary vacuum one for the given model\cite{Shar2}. This new ground state
can be regarded as the true vacuum and it exhibits the symmetry breaking.

    Before we present our results, let us describe Ghoshal and
Chatterjee's model and results briefly\cite{Shar2}.
The Hamiltonian of the model of Ghoshal and Chatterjee is as
follows\cite{Shar2}
\be
 H=\om_aa^+a+\om_bb^+b+\ka(a^+b^++ab+a^+b+b^+a)
\ee
where $a^+(a)$ is the creation(annihilation) operator for an optical
phonon of frequency $\om_a$, $b^+(b)$ is the corresponding operator for
the photon field, $\om_b$ being the photon frequency and $\ka$ is the
phonon-photon coupling strength. Using the following transformation
\be
  \left\{\ba{l}
	      a=A_1\al+A_2\al^++B_1\bt+B_2\bt^+\\
	      b=B_3\al+B_4\al^++A_3\bt+A_4\bt^+
     \ea
 \right.
\ee
and choosing suitable parameters $\{A_i,\, B_i\}$ ($i=1,\,2,\,3,\,4)$,
the expression of Hamiltonian can be diagonalized as\cite{Shar2}
\be
H=E_\al\al^+\al+E_\bt\bt^+\bt+E_0
\ee
where $E_\al$, $E_\bt$, and $E_0$ depend on $\om_a$, $\om_b$ and
$\ka$\cite{Shar2}.

  We take the form of solution as
\be
  |A\ra=e^{\al a^+a^++\bt b^+b^++\ga a^+b^+}|0\ra
\ee
where $\al$, $\bt$, and $\ga$ are parameters to be determined, $|0\ra$ is the
ordinary vacuum of the phonon and photon. If this state is really a solution
of the model, it should obey the Schr\"{o}dinger equation
\be
   H|A\ra=E|A\ra
\ee
Substituting Eq.(1) and Eq.(4) into Eq.(5), then comparing the terms of
$|A\ra$, $a^+a^+|A\ra$, $b^+b^+|A\ra$ and $a^+b^+|A\ra$ respectively, we obtain
\bea
 && E=\ka\ga                                                  \\
 && \om_a\al+\ka\ga+\ka\ga\al=0                               \\
 && \om_b\bt+\ka\ga+\ka\ga\bt=0                               \\
 &&\om_a\ga+\om_b\ga+\ka+\ka\bt+\ka\al+\ka(\al\bt+\ga^2)=0
\eea
>From these equations the equation satisfied by $E$ is obtained
\be
 (E+\om_a)(E+\om_b)-\om_a\om_b+{\ka^2\om_a\om_b\over(E+\om_a)(E+\om_b)}=0
\ee
Introducing
\be
  S=(E+\om_a)(E+\om_b)
\ee
Eq.(10) can be rewritten as
\be
  S^2-\om_a\om_b S+\ka^2\om_a\om_b=0
\ee
We have
\be
  S={\om_a\om_b\pm\sq{\om_a\om_b(\om_a\om_b-4\ka^2)}\over 2}
\ee
Then four solutions are obtained
\be
  E={1\over 2}[-(\om_a+\om_b)\pm\sq{\om_a^2+\om_b^2
     \pm2\sq{\om_a\om_b(\om_a\om_b-4\ka^2)}}]
\ee
if
\be
\om_a\om_b>4\ka^2
\ee
When we impose the normalization condition to these wavevectors, we have
the following constraints for $\alpha$, $\beta$ and $\gamma$
\be
 \left\{\ba{l}
	\alpha\!<\!\frac{1}{2}\\
	\beta\!<\!\frac{1}{2}   \\
	\gamma\!<\!1
       \ea
 \right.
\ee
With these constraints and the condition Eq.(15), only one solution
is left, {\it i.e.},
\be
E={1\over 2}[-(\om_a+\om_b)+\sq{\om_a^2+\om_b^2
    +2\sq{\om_a\om_b(\om_a\om_b-4\ka^2)}}]
\ee
This energy is negative and lower than the ordinary vacuum energy. Therefore
there is only one true vacuum. This ground state does not display the symmetry
of the Hamiltonian under the gauge transformation
\be
  U=exp\{i(a^+a+b^+b)\}
\ee
Then we have a broken symmetry.


   Let us discuss briefly the physical meanings of the new vacuum state.
If its phonon frequency( $\omega_q$ ) is approximately independent of the
wavevector ( $q$ ), the material can be regarded as the one with only one mode
of phonon. We put this material into a cavity
with one mode of radiation field. When the temperature gets lower and
lower, the probability of the new vacuum increases larger and larger and we will
find even the temperature approaches to the absolute zero, the
average number of phonons arrives at a definite non-zero value. This phenomenon
is quite different from the one without the existence of the new vacuum state.
If there would not be the new vacuum,  the ordinary vacuum $|0\ra$ would
dominate in the probability distribution when the temperature approaches
absolute zero, i.e., the phonon number in the material would tend to be
zero.


  This new vacuum state will have influence on some physical quantities which
could be measured by the experiment. When the material
is put in the cavity under enough low temperature there will be transitions
from the exited states of polariton to the new vacuum state, and then
some new lines in spectrum will be found. For the same reason, there will be
new structures in the absorption spectrum. On the other hand, this
new vacuum state may also have the contribution on the specific heat of the
material. We hope these interesting results can stimulate experiments to
study the polariton system.

    From the discussions above, we can conclude here, that in the model
polariton system we studied, there is a true vacuum state, which has really
lower energy than $|0\ra$, provided the condition Eq.(15) is satisfied.
This result is interesting because the true ground state or the symmetry
breaking is obtained without any assumption and approximation in this case.

\section*{ACKNOWLEDGMENTS}
   This work was supported by the  National Climb Project ``Nonlinear
Science'' of China. We would like to thank Professor S.L. Wan for helpful
discussions.

\end{document}